%% file: aaskaii_template.tex
\documentclass[a4paper,11pt]{article}
\usepackage{aaskaiid}
\usepackage{orcidlink}

\title{Resolved H\textsc{i} and Environmental Dynamics}

\author[1,2]{M.~Ramatsoku\orcidlink{0000-0003-0231-3249}}
\ShortName{M.~Ramatsoku et al.}

\author[2]{P.~Serra\orcidlink{0000-0001-5965-252X}}
\author[3]{N.~Deg\orcidlink{0000-0003-3523-7633}}
\author[4]{R.~Ianjamasimanana\orcidlink{0000-0003-2476-3072}}
\author[4]{A.~Sorgho\orcidlink{0000-0002-5233-8260}}
\author[5,6]{G.~De~Lucia\orcidlink{0000-0002-6220-9104}}
\author[3]{K.~Spekkens\orcidlink{0000-0002-0956-7949}}
\author[4]{L.~Verdes-Montenegro\orcidlink{0000-0003-0156-6180}}
\author[7,8]{H.~Yoon\orcidlink{0000-0003-4048-2203}}
\author[4]{B.~Namumba\orcidlink{0000-0003-1032-8889}}
\author[9]{M.~Meyer\orcidlink{0000-0002-2838-3010}}

\affiliation[1]{Centre for Radio Astronomy Techniques and Technologies (RATT), Department of Physics and Electronics, Rhodes University, Makhanda 6140, South Africa}

\affiliation[2]{INAF – Osservatorio Astronomico di Cagliari, Via della Scienza 5, I-09047 Selargius (CA), Italy}
\emailAdd{m.ramatsoku@ru.ac.za}

\affiliation[3]{Department of Physics, Engineering Physics, and Astronomy, Queen’s University, Kingston ON K7L 3N6, Canada}
\affiliation[3]{Arthur B. McDonald Canadian Astroparticle Physics Research Institute, Queen’s University, Kingston ON K7L 3N6, Canada}
\emailAdd{nathan.j.deg@gmail.com}

\affiliation[4]{Instituto de Astrofísica de Andalucía (IAA-CSIC), Glorieta de la Astronomía s/n, 18008 Granada, Spain}
\emailAdd{ianja@iaa.es}
\emailAdd{asorgho@iaa.es}

\affiliation[5]{INAF – Astronomical Observatory of Trieste, Via G. B. Tiepolo 11, 34143 Trieste, Italy}
\affiliation[6]{IFPU – Institute for Fundamental Physics of the Universe, Via Beirut 2, 34151 Trieste, Italy}

\affiliation[7]{Institute for Data Innovation in Science, Seoul National University, 1 Gwanak-ro, Gwanak-gu, Seoul 08826, Republic of Korea}
\affiliation[8]{Astronomy Program, Department of Physics and Astronomy, Seoul National University, 1 Gwanak-ro, Gwanak-gu, Seoul 08826, Republic of Korea}

\affiliation[9]{International Centre for Radio Astronomy Research (ICRAR), The University of Western Australia, 35 Stirling Highway, Crawley WA 6009, Australia}

\abstract{Spatially resolved, deep \HI\ observations from SKA precursors and pathfinders such as MeerKAT, FAST, and ASKAP have demonstrated their ability to reveal the complex interactions between galaxies and their environments. These include, but are not limited to, recent observations of the Virgo cluster showing that the hydrodynamical effects of ram pressure stripping can operate effectively at unexpectedly large cluster-centric distances. In the Fornax cluster, the discovery of long \HI\ tails with mixed tidal-ram-pressure origins indicates the interplay between gravitational and hydrodynamical mechanisms. Similar \HI\ features in nearby filaments and galaxy groups, where ram pressure is expected to be weak, highlight the influence of hydrodynamical processes even in low-density environments. Multi-resolution studies have further revealed signs of cold gas accretion and \HI\ replenishment driven by tidal interactions.
While highly informative, these studies remain limited to small, specific regions of the sky. With SKA-mid AA4, it will become possible to carry out deep, spatially resolved \HI\ imaging over hundreds of square degrees, covering environments from isolated galaxies to filaments. By reaching column-density sensitivities between $1.0 \times 10^{18}$ and $\sim 1.0 \times 10^{19}~\mathrm{cm^{-2}}$ at physical resolutions of $\sim$10 and $\sim$1 - 2 kpc, respectively, and by enabling sensitive, contiguous observations of wide areas within short integrations, SKA-mid AA4 will allow the construction of large, statistically representative samples of galaxies and detailed studies of environmental mechanisms operating across the full range of these less-studied environments at resolved scales.}

\input{abb}

\begin{document} 
\maketitle

\section{Introduction}

A well-established trend observed in studies of galaxy evolution is that the number of passive, early-type galaxies increases with rising galaxy density, while the fraction of actively star-forming, late-type galaxies decreases (e.g., \citealp{DresslerMorphClust1980}). The neutral atomic hydrogen (\HI) provides a fundamental probe into this transformation. As the dominant component of the cold interstellar medium (ISM) and the primary reservoir from which molecular gas and new stars form, \HI\ is essential for sustaining ongoing star formation. Any changes in its content or spatial distribution can have a profound influence on the evolutionary pathway of galaxies, making \HI\ an especially sensitive tracer of the physical processes that regulate galaxy evolution across environments. The exact physical mechanisms responsible for these transformations remain incompletely understood, but are broadly thought to originate from interactions between galaxies and their surroundings.
Many of these mechanisms can affect the star formation activity of galaxies by disrupting or removing their gas, ultimately changing them from active star-forming to passive quenched systems. 

One such process is starvation, whereby the cooling and accretion of the gas from the circumgalactic medium (CGM) onto the disc is prevented, leading to a slow consumption of the cold gas and decline of the star formation rate \citep{Larson1980}. More intense galaxy-environment interactions directly disturb or remove the cold interstellar medium (ISM) within the disc. These include gravitational interactions such as galaxy-galaxy collisions, ranging from low-velocity encounters, which may lead to mergers, to high-velocity flybys, which can disturb both the stellar component and the ISM, thereby removing or consuming the available fuel for new star-formation, (e.g. \citealp{Toomre1972, Farouki1981, Moore1996, Moore1998}).

In other cases, the ISM of galaxies interacts with the intergalactic medium (IGM) embedded in the cosmic web. Such interactions can displace or even completely remove the ISM through ram pressure \citep{Gunn1972} and/or viscous stripping \citep{NulsenViscStrip1982}, and thermal evaporation \citep{CowieSongailaEvap1977}. These mechanisms can have complex short-term effects on the star-formation activity of galaxies, such as temporary enhancements through compression of the ISM \citep{Bekki2003} or chaotic cold accretion (CCA) driven by increased gas turbulence \citep{Gaspari2013}. Ultimately, however, these interactions remove or exhaust the gas supply necessary for ongoing star formation, leading to quenching (see \citealp{Cortese2021, Boselli2022} for full reviews). 

The relative roles and characteristic timescales of these mechanisms are reasonably well-constrained for low-redshift clusters and their immediate surroundings. However, most galaxies in the local Universe reside outside clusters, in lower-density regions such as filaments and groups, which are known to feed galaxies and into clusters along the cosmic web \citep{Cautun2014}. In these environments, the picture is significantly less well-established: hydrodynamical effects are expected to be weaker, while gravitational encounters may play a more prominent role due to lower galaxy velocities and the reduced density of the IGM (e.g. \citealp{Das2023}). Observational information, however, remains limited. 
For instance, several studies suggest that a galaxy’s distance from the filament spine correlates with its gas content, star-formation activity, and stellar mass. Galaxies located closer to filament spines tend to have higher average stellar masses \citep{Kraljic2018} and to experience interactions more frequently \citep{SantiagoBautistaFilaments2020}, which can accelerate the quenching of star formation \citep{LaigleFilaments2018}. However, conflicting results have been reported. For example, \citet{KuutmaFIlaments2017} did not find evidence for increasing stellar mass towards filament spines, although they observed higher early-type fractions and lower star-formation rates closer to filaments. Recent work has also questioned whether filaments themselves exert a strong direct influence on galaxy evolution. Using large statistical samples and carefully constructed control populations, \citet{Zakharova2023} argued that the majority of the observed trends in gas content, morphology, and star-formation activity near filament spines are driven primarily by the prevalence of galaxy groups embedded within filaments rather than by the filaments themselves. In this view, filaments serve mainly as conduits funnelling galaxies into group environments, where most environmentally driven gas removal and transformation occur.
Furthermore, earlier studies based on unresolved \HI\ data from large-area surveys such as HIPASS \citep{BarnesHIPASS2001} and ALFALFA \citep{GiovanelliALFALFA2005} have reported that galaxies located in proximity to filament spines often exhibit reduced \HI\ content \citep{OdekonFilaments2018, HoosainFilaments2024}. However, the limited sample sizes in these works have led to conflicting results regarding the stellar mass regime over which this trend holds. Notably, \citet{KleinerCosmicWeb2017} observed an opposite behaviour; an excess of \HI\ in the most massive galaxies. Because these surveys provide only integrated \HI\ measurements, they lack the spatial resolution necessary to examine how environmental mechanisms shape the internal distribution and kinematics of the ISM.

\subsection{Resolved \HI\ - A powerful tracer of mechanisms operating in the cosmic web}
\vskip -5pt
The use of sensitive and resolved \HI\ imaging has proven to be a powerful tool to study the environmental interactions governing galaxy evolution. \HI\ extends out to large galaxy radii at the interface between galaxies and their environment, making it one of the first components to respond to external mechanisms. In these outer regions of galaxies, the gravitational bond with the host galaxy is relatively weak and the gas is more diffuse, making the \HI\ gas susceptible to perturbations and enabling environmental processes to leave distinct imprints on the diffuse \HI\ disc. Moreover, the long dynamical timescales in these outskirts (on the order of $\sim$1~Gyr) allow \HI\ to retain signatures of past interactions over long periods. Studying these extended \HI\ features in galaxies' outskirts requires a combination of resolution and sensitivity.

\HI\ imaging maps from the SKA L-band precursors and pathfinders, JVLA, MeerKAT, FAST, ASKAP and others, has proven invaluable for revealing the complexity of galaxy-environmental interactions. An increasingly large population of galaxies is now being identified with unusual morphologies, including jellyfish galaxies, tidal tails and bridges, extreme warps, polar rings and more. While it is well established that these features arise from galaxy–environment interactions, the specific mechanisms driving their formation can vary considerably. Thus, studying galaxies with unusual morphologies can help identify and quantify the environmental processes at play.

\begin{figure}
    \centering
    \includegraphics[width=0.55\linewidth]{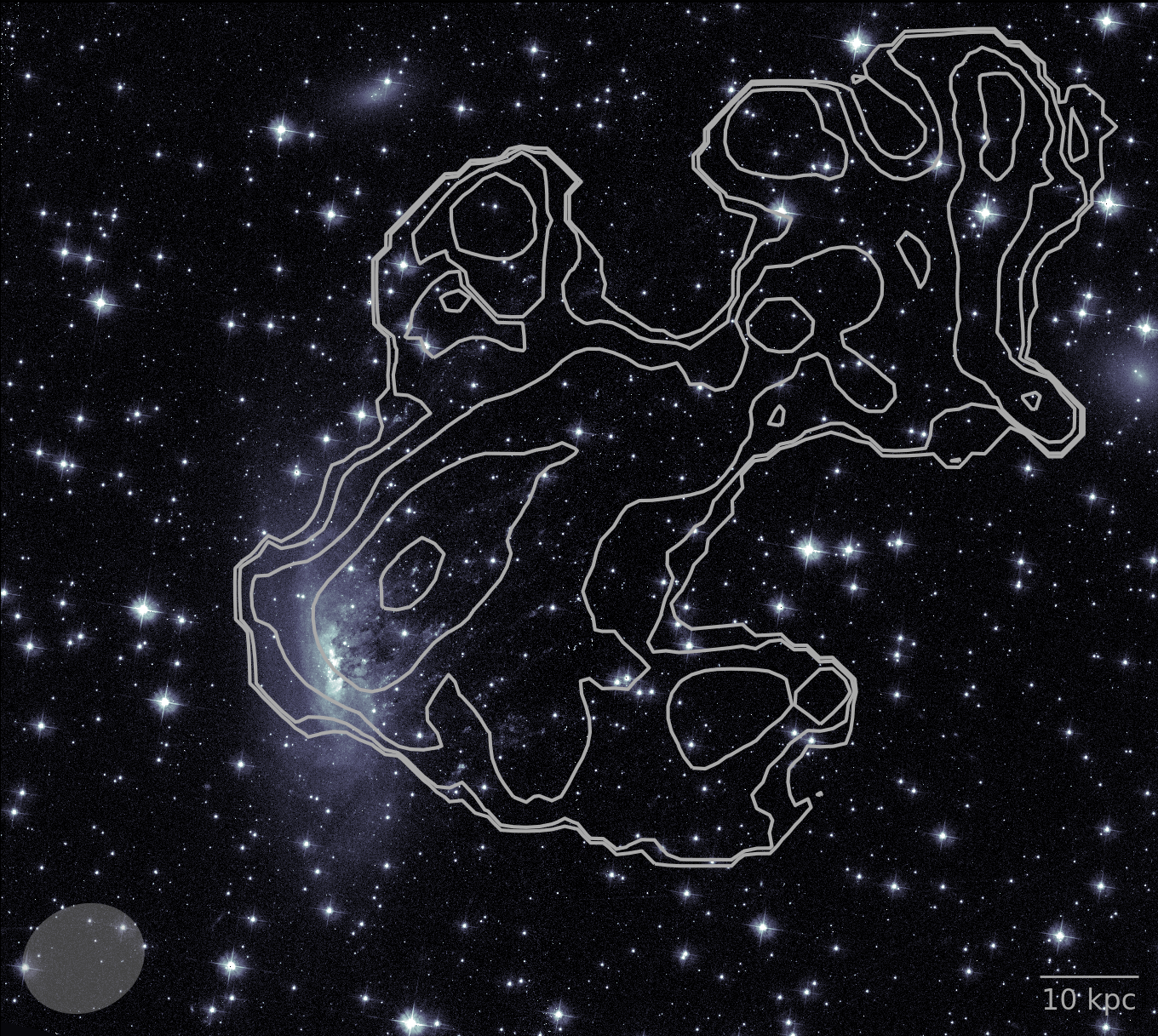}
    \includegraphics[width=0.37\linewidth]{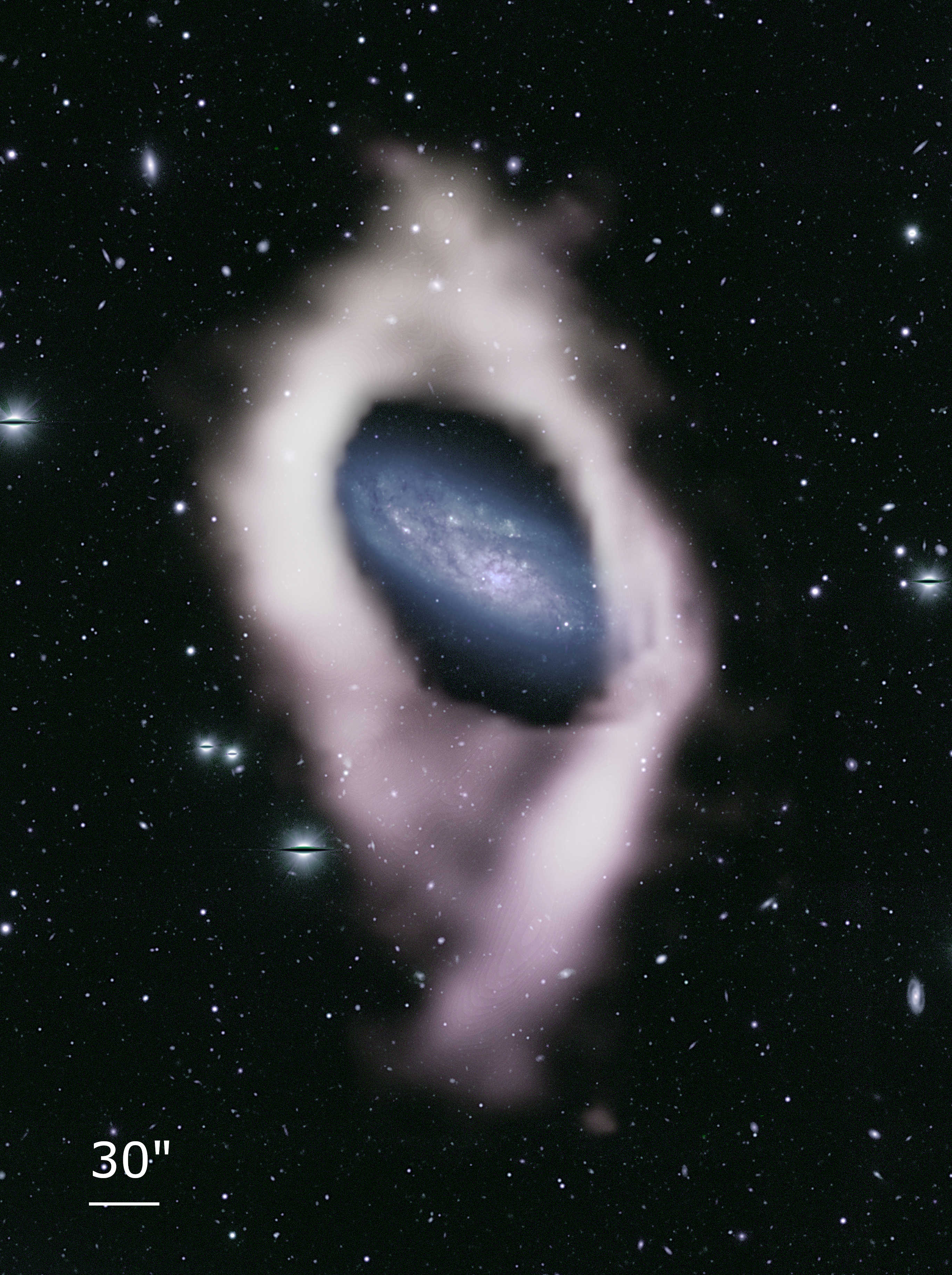}

    \caption{Examples of unusual H\textsc{i} morphologies reported in recent studies. \emph{Left panel}; the H\textsc{i} spatial distribution of the jellyfish galaxy ESO137-001 is shown overlaid on the optical map and the HST WFC3 image. The H\textsc{i} contours drawn at column densities of \(N_{\mathrm{HI}} = 2.3,\ 4.6,\ 9.2,\ 18.4 \times 10^{19}\ \mathrm{atoms\,cm^{-2}}\), clearly traces the characteristic jellyfish morphology previously observed in the optical and at other wavelengths \citep{Ramatsoku2025}.
  \emph{Right panel}; the anomalous H\textsc{i} component of the polar ring galaxy NGC 4632, illustrating that this gas is clearly oriented differently from the stellar disc.  The gas in the main disc has been removed to highlight the polar structure \citep{Deg2023}. Images are reproduced with permission.}
    \label{fig:polar_ring}
\end{figure}

Jellyfish galaxies exhibiting one-sided \HI\ tail are thought to result from the aforementioned ram-pressure stripping as galaxies move through dense environments (see example in the left panel of Fig.~\ref{fig:polar_ring}). Strong ram pressure can remove \HI\ gas and produce highly asymmetric morphologies \citep[e.g.,][]{Boselli2006, Fumagalli2014, McPartland2016}, while compression of the interstellar medium may form knots of young stars in the stripped tails \citep[e.g.,][]{Yoshida2008, Poggianti2019} while also triggering enhanced star formation \citep{Vulcani2018, Ramatsoku2019}. Although initially identified in massive clusters \citep{Yagi2007,Chung2007, Chung2009,Poggianti2016}, recent observations have revealed jellyfish-like morphologies in lower-mass groups and even filaments, showing that they can occur across a wide range of environments \citep[e.g.,][]{Roberts2021}.

Even more complex morphologies have been reported in recent observations of the Virgo cluster, which revealed an extended \HI\ tail in a galaxy located about 3$\deg$ (approximately 0.86~Mpc) from the cluster centre, similar to those seen in systems undergoing hydrodynamical interactions with the intracluster medium (ICM) in cluster cores \citep{Boselli_pilot_vic}. 
In the low-mass Fornax cluster, the discovery of long \HI\ tails with a mixed tidal-then-ram-pressure \citep[e.g.,][]{Serra2023, serra2024} origin suggests a complex balance between these mechanisms. Similar \HI\ features have also been observed in deep \HI\ images of the galaxy groups in the MeerChoir survey and filaments like those found in the  Virgo III filament located $\sim 5$ Mpc to the south-east of the Virgo cluster core \citep{Finn2025}. Moreover, a recent study \citet{Staveley-Smith2025} reported on the tidal bridge detected by WALLABY, which also exhibits a significantly longer tidal tail. Although this system lies outside the virial radius of the Virgo cluster, simulations indicate that the envelope of hot gas has shaped the tidal tail, and the ongoing interaction with the cluster has prevented a rapid merger \citep{Westmeier_wallaby,Murugeshan2024}.  

Polar ring galaxies are another striking phenomenon, where a host galaxy contains material orbiting at $\sim90^{\circ}$ to the main body, an example of which is shown in the right panel of Fig.,~\ref{fig:polar_ring}.  Typically, these have been found in the optical (see, for example, \citealt{Schecter1978,Whitmore1990,Moiseev2011}), however, many have also been detected in \hi\ \citep{vanGorkom87,Brook2008,Dzudzar2021}.  With the advent of resolved, sensitive surveys on SKA pathfinders, more polar ring galaxies are being detected in \hi\ \citep{Deg2023,Healy2024}. For example in \citet{Deg2023} galaxies with multiple \hi\ components in the first WALLABY pilot data release \citep{Westmeier_wallaby} were reported. The anomalous gas in those systems is consistent with being polar.  And \citet{Healy2024} posited that a polar structure may explain the anomalous gas revealed by the MHONGOOSE survey. In another study \citet{Stanonik2009}, a \HI\ polar-ring was detected in a wall between two voids with an \HI\ mass comparable to the stellar mass and hints of a warp, indicating slow and recent gas accretion. The blue, UV-bright central disk detected in this and an underdense environment supports cold-flow accretion as a viable formation mechanism \citep{Stanonik2009}.
\citet{Deg2023} went further and estimated a possible incidence rate for polar ring galaxies of $1-3\%$, which is an order of magnitude larger than the previously estimated optical rate of $0.1\%$ \citep{Reshetnikov2011}.

An alternate explanation for these systems is that they are `extreme' warps.  In such systems, the material may not be orbiting at $90^{\circ}$ and is connected to the main body of the galaxy. Regardless of the precise classification, both polar ring galaxies and extreme warps must be formed by interaction events. These types of interactions include mergers, fly-by passages that deposit material on the host galaxy, accretion from the cosmic web, and more.  Building up a statistical sample of well-resolved polar rings and extreme warps will constrain the precise origin mechanism, help to distinguish between the differing classes, and explore how the stability of these features affects their host galaxies.

Through these resolved and sensitive \HI\ studies, a complex picture of the interplay between gravitational and hydrodynamical interactions is starting to emerge but they focus on small specific regions. The picture is made more complex by the recent result of tidal interactions possibly triggering the accretion of cold gas in a few individual objects (\citealp{Wang2023}) showing that detailed observations of larger samples are critical. These systems also provide excellent testbeds for studies for how pre-processing modifies systems as they fall into denser environments \citep{Fujita2004}.


Clarifying the formation pathways of these unusual morphologies is therefore essential for constraining the physical processes that drive environmental interactions and for understanding their role in galaxy evolution. Such investigations demand the high resolution, sensitivity, and survey speed which can be achieved with the SKA-mid-AA4, which will make it possible to build statistically significant samples of galaxies with resolved \HI\ morphologies across diverse environments. 


This has been a major challenge for existing and planned surveys, as none of them simultaneously meet all the observational requirements. The currently ongoing and previous \HI\ imaging surveys lack the necessary combination of sensitivity, spatial resolution, and sky coverage required to encompass the full range of environments and to resolve the detailed structure of galaxy discs and their outskirts.

\begin{figure}
    \centering
\includegraphics[width=0.8\linewidth]{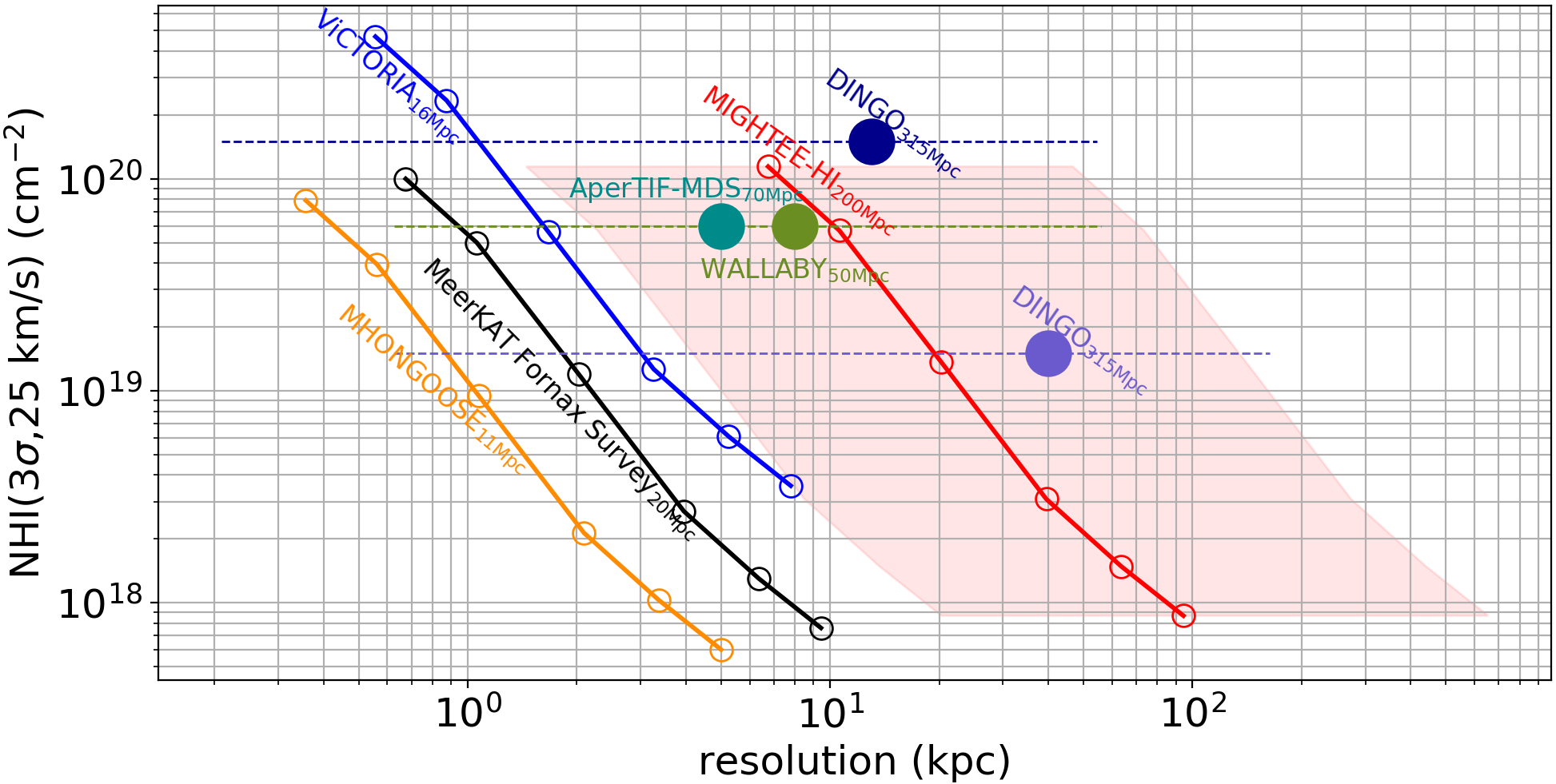}
\caption{Comparison of the typical \HI\ column-density sensitivities achieved by existing \HI\ imaging surveys as a function of physical resolution, evaluated at the representative redshift or at the redshift of the main large-scale structure targeted. These redshifts are indicated with text annotations next to each line. For the large-area surveys we also show the sensitivity and physical resolution across their redshift range, e.g.,  red shaded area for MIGHTEE and dotted lines for WALLABY and DINGO  (see also \citealt{maccagni2024} for additional examples). LADUMA and CHILES are not shown due to their too-high redshifts.}\label{fig:existingsurveys}
\end{figure}

A few representative examples are shown in Fig.\,\ref{fig:existingsurveys} to illustrate this limitation. Surveys that cover large volumes, such as LADUMA \citep{Blyth_laduma}, CHILES \citep{Fernandez_chiles}, and MIGHTEE-HI \citep{Maddox_mighteeHI}, lack the sensitivity required to trace the extended \HI\ features of galaxies at the physical resolution of a few kpc, making them unsuitable for studying environmental processes at resolved scales. 
Low-redshift surveys such as WALLABY \citep{Westmeier_wallaby}, DINGO \citep{Meyer_dingo}, Apertif medium and shallow surveys \citep{Adams_apertif} will detect thousands of galaxies in \HI\ and offer large sample sizes, enabling the study of global \HI\ properties in different environments, but lack sufficient sensitivity and physical resolution.
Other surveys, including the MeerKAT Fornax Survey (MFS; \citealt{Serra2023}), ViCTORIA \citep{Boselli_pilot_vic}, and MHONGOOSE \citep{deBlok2024_mhongoose}, achieve excellent sensitivity and provide the few-kpc resolution necessary to investigate instabilities in galaxy discs and extended \HI. However, the first two focus only on specific environments, Fornax and Virgo clusters, and therefore do not sample the full range of environments, while the latter targets individual, very nearby galaxies and consequently does not cover other large-scale structures.

\section{Physical processes across environments}
Combining resolved \HI\ maps with stellar data from optical imaging will reveal the aforementioned characteristic morphological signatures as a function of environment. A key challenge faced in large, untargetted surveys, is identifying targets of interest.  A tool that may potentially help are statistical measures of morphology \citep{Conselice2000,Conselice2003,Lotz2004}. These so-called `morphometrics' have been widely used in optical studies \citep{Conselice2003,Lotz2004,Pearson2019,Bellhouse2022,Zhao2022,Sazonova2024}.  In \hi\ studies, the parameter most commonly used in older studies is the profile asymmetry \citep{Peterson1974,Espada2011,Bok2019,Deg2020,Reynolds2020}. However, \citet{Holwerda2011, Giese2016} and \citet{Holwerda2025} used the full suite of morphometrics to study THINGS galaxies \citep{Walter2008}, WHISP galaxies, and WALLABY pilot observations respectively.
Recently, \citet{Deg2023b} developed a 3D asymmetry morphological statistic for use with \hi\ observations that appears to be applicable at lower radii than comparable 2D asymmetry morphometrics. \citet{PerronCormier2025} used this statistic on both simulations and WALLABY observations. Not only did the 3D statistic reliably detect galaxies with disturbed and unusual morphologies, it was possible to do a first comparison of simulated morphologies to observations on a statistical basis.  This result highlights the dual promises of morphometrics; they can indeed find the unusual morphologies, and can be used as a test for cosmological simulations and constrain galaxy formation.  Moreover, it is possible to use the population of disturbed galaxies as a proxy to measure the merger rate across a variety of environments which is an analysis that will become routine with the sensitivity and speed of the SKA.  

\subsection{Quantifying the effects of environment on galaxy kinematics} 
Early large-sample studies such as WHISP \citep{Swaters2002} and higher-resolution mapping efforts such as THINGS \citep{Walter2008} and LITTLE THINGS \citep{Hunter2012} revealed a variety of non-axisymmetric features and warped velocity fields; however, the relatively small and heterogeneous nature of these samples make it hard to draw statistical conclusions. Consequently, it remains unclear whether disturbed kinematics are exceptional or rather an intrinsic property of gas discs once a broad range of masses and environments is probed. Further ambiguity arises from internal drivers of non-circular motions. Observational and numerical studies indicate that stellar feedback can drive turbulence within the neutral interstellar medium, leading to broader \HI\ line profiles \citep[e.g.,][]{Iorio2017}. Using the aforementioned 3D asymmetry morphological statistic, it is demonstrated that part of the observed kinematic complexity may be intrinsic, rather than exclusively caused by tidal interactions or environmental influences. At the same time, cosmological hydrodynamical simulations from the NIHAO and FIRE projects emphasise a large diversity of \HI\ kinematic morphologies at fixed halo mass \citep[e.g.][]{WangL2015, Hopkins2020}, further complicating observational interpretation. Untargetted \HI\ surveys of galaxies across diverse environments at kpc-resolution and high spectral fidelity will provide a statistically robust and spatially resolved view of the frequency and nature of disturbed \HI\ kinematics, helping to determine whether such features are the rule or the exception, and to what extent stellar feedback contributes to them.

\subsection{Gas contents across different environments}
\begin{figure}
    \centering
\includegraphics[width=0.8\linewidth]{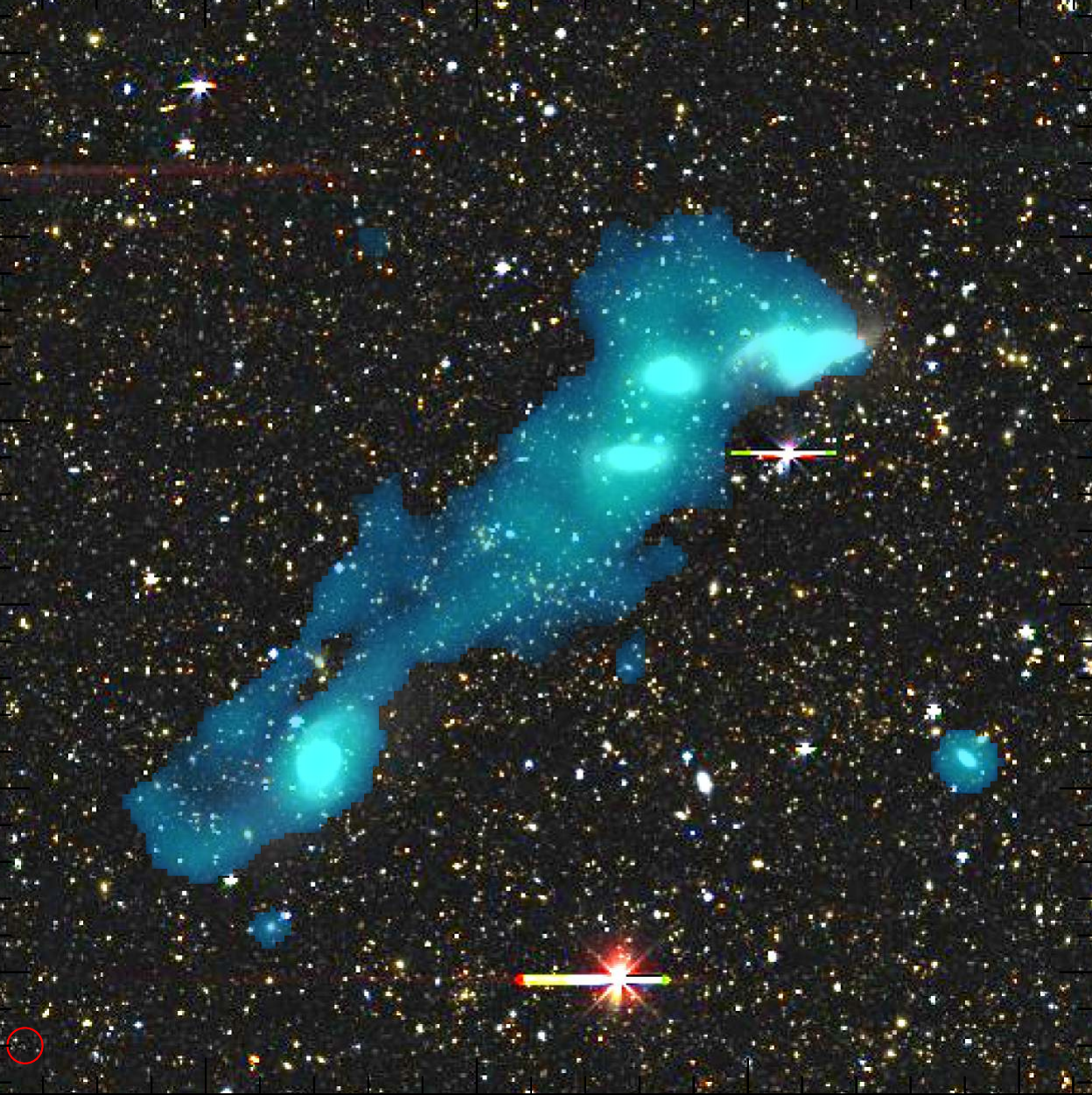}
\caption{The \HI\ distribution of HCG 16 as seen by MeerKAT  \citep{Ianja2025,Sorgho2025}. The blue hue represents the \HI\ emission from the group at a threshold column density of $3.5\times10^{18}\rm\,cm^{-2}$. The synthesised beam of $58''$ ($\rm{\sim}14\,kpc$) is shown at the bottom-left corner in red. The background image shows DECaLS optical data to highlight the core members. Images reproduced with permission from the authors.}\label{fig:hcgs}
\end{figure}

Although most galaxies reside in groups, these environments remain less well studied than clusters. Groups range from loose associations to compact systems, with the latter showing accelerated evolution driven by rapid gas removal. Hickson Compact Groups \citep[HCGs;][]{Hickson1982}, consisting of four to ten galaxies in dense configurations, display a wide variety of \HI\ and optical morphologies \citep{Huchtmeier1997, Verdes2001} and therefore serve as excellent laboratories for studying environmental interactions. An evolutionary sequence of the HCGs, based on their \HI\ morphologies, was proposed over two decades ago \citep{Verdes2001} and was refined by \citet[e.g.,][]{Jones2023}. In this scheme, groups are assigned an \HI\ phase according to how much of the detected \HI\ remains directly associated with the galaxies versus being found in extended intra-group features (tails, bridges, diffuse structures). In phase~1, galaxies keep most of their detected \HI\ within the disc. In phase~2, between 25\% and 75\% of the detected \HI\ is found in extended features, and phase~3 systems are either \HI-undetected or have more than 75\% of their detected \HI\ in extended structures \citep{Verdes2001,Jones2023}. Recent efforts conducted with VLA and MeerKAT established a broad picture of the large-scale neutral gas distribution in HCGs beyond any previous observational results \citep{Jones2023,Ianja2025,Sorgho2025}. Consistent with the aforementioned evolutionary model, this picture tentatively shows that late-phase (phase 3) HCGs are significantly distinct from their immediate environments, behaving as separate entities located in regions that otherwise have different gas contents. However, when observed in the early or intermediate stages of their life, HCGs do not show systematic differences with respect to their environments in terms of gas content \citep{Sorgho2025}. SKA1-Mid AA4 will be able to test this picture decisively by mapping \HI\ in and around HCGs with uniform sensitivity and resolution across statistically meaningful samples. With a $\sim$6 h L-band track, previous MeerKAT observations \citep{Ianja2025,Sorgho2025} failed to detect diffuse \HI\ emission in late-phase HCGs that were apparent in single dish observations \citep{Borthakur2010}. Over similar integration times, the SKA's AA4 array will be able to reach column-density limits of the order of $10^{18}\ \mathrm{cm}^{-2}$ over ${\sim}20\ \mathrm{km\,s^{-1}}$, about 3 times more sensitive than MeerKAT. This will allow us to detect (or place stringent limits on) the diffuse intra-group reservoirs that govern the \HI\ budget of the late-phase HCGs. The MeerKAT observations by \citet{Ianja2025,Sorgho2025} revealed intricate \HI\ tidal tails in intermediate-phase HCGs that were previously missed by the VLA (see Figure~\ref{fig:hcgs}). SKA AA4 will be able to trace the full extent of \HI\ in these HCGs, allowing us to reassess their \HI\ content and deficiency parameters.

At the opposite end of the spectrum, isolated galaxies represent a benchmark for estimating the impact of the environment on the \HI\ content of galaxies. Over decades, effort was put into defining a nurture-free sample of galaxies, AMIGA \citep[Analysis of the interstellar Medium in Isolated GAlaxies][]{Verdes2005a}, on which environmental effects are minimal. 
Thanks to their high degree of isolation, AMIGA-like galaxies are ideal targets for the search for signs of cold gas accretion. This process of gas replenishment in present-day galaxies is predicted to consist of filaments of cool gas entering a galaxy's halo to feed its disc \citep{Keres2005}. Over the past several years, several studies have been undertaken to detect direct evidence of gas accretion in nearby galaxies. The HALOGAS survey, which mapped the \HI\ distribution in 22 galaxies (down to ${\sim}10^{19}\rm\,cm^{-2}$) with the WSRT, identified several \HI\ clouds and streams in the vicinity of the sample galaxies whose origin could be attributed to accretion processes \citep{Heald2011,Heald2015,Kamphuis2022}. More recently, the MHONGOOSE project provided deep, high-resolution \HI\ data of 30 nearby, gas-rich spiral and dwarf galaxies observed with MeerKAT \citep{deBlok2024_mhongoose}. Although conclusive evidences of gas accretion have yet to be found, ongoing efforts are promising. Leveraging the resolution and sensitivity capabilities of SKA-Mid AA4 to focus the search on extremely isolated galaxies, such as those of AMIGA, will provide irrefutable proof or constraints on cold gas accretion in the local universe.

\section{SKA-mid and resolved H\textsc{i}-environment dynamics}
Over the past decades, observations with SKA precursors (\citealp{serra2024, deBlok2024_mhongoose, Healy2024,  deGasperin2025}) together with cosmological simulations (\citealp{Popping2009, Goller2023}) have demonstrated that column-density sensitivities of at least $\sim 10^{19}~\mathrm{atoms\,cm^{-2}}$ are required to detect key signatures of environmental processes in the diffuse \HI\ phase of the intergalactic medium. For nearby systems surveyed at kpc-scale resolution (e.g.\ MFS, MHONGOOSE, VICTORIA), diffuse \HI\ tails and star-forming clumps are routinely detected at $\sim 10^{19}~\mathrm{atoms\,cm^{-2}}$ at 1 -- 2 kpc resolution and $\sim 10^{18}~\mathrm{atoms\,cm^{-2}}$ at $\sim 10$ kpc over integration times of 10 -- 50 hours (\citealp{Serra2023, deBlok2024_mhongoose, Boselli_pilot_vic}). These results strongly motivate an AA4–SKA-mid survey designed to blindly image a large contiguous sky area of some hundreds deg$^2$ comprising voids, groups, and filaments, enabling a statistically significant galaxy sample with similar sensitivities and physical resolutions.

According to the SKA-Mid sensitivity calculator (March 2026, version 2.4.1), within Band~2 (0.95 -- 1.76~GHz) of AA4, and assuming all antennas are available, the desired column density sensitivity can be achieved with a rms noise level of $\mathrm{rms} \approx 112~\mu\mathrm{Jy~beam^{-1}}$ per 26.9 kHz-wide channel. Assuming a Briggs weighting of $r = 0$ and a synthesised beam with a FWHM of $10\arcsec \times 10\arcsec$, this sensitivity corresponds to an integration time of approximately 7~hours over a spectral resolution of 5.7 \kms.

With these parameters, we expect column density sensitivities of approximately $\sim 4 \times 10^{19}~\mathrm{cm^{-2}}$ at $3\sigma$ assuming a linewidth of 25~\kms. To achieve the desired physical resolution, observations should target galaxies at distances of roughly 40~Mpc, which is the optimal range for achieving a spatial resolution of $\sim$1--2~kpc. To encompass all necessary environments a mosaic covering about 200 -- 250~deg$^{2}$ would be required. Achieving this sensitivity requires an effective observing time of $\sim13$~hours per square degree, for a total of approximately 2600~hours. This estimate incorporates the primary beam response and the net sensitivity gain from a Nyquist sampled mosaic, ensuring uniform depth across the survey area and enabling the detection of diffuse \HI\ at kpc-scale physical resolutions in a wide range of environments. This would enable studies such as those of the MeerKAT Fornax or the VICTORIA survey (\citealp{Serra2023, Kleiner2023}), to be  at scale. 

For comparison, achieving the same sensitivity and physical resolution with current facilities such as MeerKAT would require approximately $60$~hours per square degree for a 200~deg$^{2}$ survey. SKA-Mid AA4 therefore, increases the survey speed by a factor of about five. This improvement will make surveys that are currently prohibitively time-consuming routinely achievable programmes.

\section{Conclusions}

Resolved \HI\ observations from SKA precursors have revealed a wide range of complex morphologies, including jellyfish tails, tidal bridges, warps, polar structures, and disturbed kinematics that trace the interplay between gravitational and hydrodynamical processes across different environments. These studies have demonstrated that well-known environmental processes operate far beyond the cluster regime where they are traditionally studied, extending into groups and filaments where their interplay is considerably more complex. However, existing observations remain confined to small, targeted regions. 
By targeting fields in the nearby Universe (within roughly 40~Mpc), kpc-scale physical resolutions can be achieved at column-density sensitivities of the order of approximately $4 \times 10^{19}~\mathrm{cm^{-2}}$ at 1 -- 2 ~kpc and $2 \times 10^{18}~\mathrm{cm^{-2}}$ at 10~kpc. These sensitivities are comparable to those obtained by current deep \HI\ surveys, but SKA-Mid~AA4 can deliver them across far larger areas due to its survey speed. It will be possible to cover a homogeneous region of approximately $200~\mathrm{deg^{2}}$ that includes less-studied environments, such as isolated galaxies, groups, and filaments, providing a large and statistically significant sample needed to quantify how environmental processes operate across the full range of these environments. 
By reducing the observing time per square degree by a factor of five relative to current pathfinders like MeerKAT, SKA-Mid AA4 makes such a survey feasible for the first time. This capability will enable the systematic identification of tidal debris, extended tails, warps, and polar structures, allowing robust measurements and the identification of gas removal processes across various environments, as well as a direct statistical comparison with cosmological simulations. It is also anticipated that one of the most advanced wide-field optical facilities, the Vera C. Rubin Observatory, will place strong constraints on stellar structure and recent star-formation activity. In synergy with high-sensitivity \HI\ imaging, which directly traces gas removal and accretion, this combination will provide a comprehensive and physically connected view of the environmental processes shaping galaxy evolution.

\bibliographystyle{abbrvnat-maxbibnames4}
\bibliography{chapter} 

\end{document}

%% file: abb.tex
\def\la{\mathrel{\hbox{\rlap{\hbox{\lower4pt\hbox{$\sim$}}}\hbox{$<$}}}}
\def\ga{\mathrel{\hbox{\rlap{\hbox{\lower4pt\hbox{$\sim$}}}\hbox{$>$}}}}

\def\arcsec{\hbox{$^{\prime\prime}$}}

\def\deg{{^\circ}}

\newcommand{\kms}{{\,km\,s$^{-1}$}}

\newcommand{\HI}{\mbox{\normalsize H\thinspace\footnotesize I}}

\newcommand{\hi}{H\thinspace{\protect\scriptsize I}}